\documentclass[12pt,fleqn]{article}
\topmargin - 0.8cm

\renewcommand{\theequation}{\arabic{section}.\arabic{equation}}
\renewcommand{\thesection}{\arabic{section}.}

\catcode`@=11
\@addtoreset{equation}{section}
\catcode`@=12
\mathchardef\SGamma="7100

\usepackage{amsfonts}
\usepackage{color}
\usepackage{amssymb}
\usepackage{amsmath}

\begin{document}
\title{\vskip-1.7cm \bf  Quantum Effective Action in Spacetimes with
Branes and Boundaries}
\date{}
\author{A.O.Barvinsky and D.V.Nesterov}
\maketitle \hspace{-8mm}{\em Theory Department, Lebedev Physics
Institute, Leninsky Prospect 53, Moscow 119991, Russia}
\begin{abstract}
We construct quantum effective action in spacetime with
branes/boundaries. This construction is based on the reduction of
the underlying Neumann type boundary value problem for the
propagator of the theory to that of the much more manageable
Dirichlet problem. In its turn, this reduction follows from the
recently suggested Neumann-Dirichlet duality which we extend beyond
the tree level approximation. In the one-loop approximation this
duality suggests that the functional determinant of the differential
operator subject to Neumann boundary conditions factorizes into the
product of its Dirichlet counterpart and the functional determinant
of a special operator on the brane -- the inverse of the
brane-to-brane propagator. As a byproduct of this relation we
suggest a new method for surface terms of the heat kernel expansion.
This method allows one to circumvent well-known difficulties in the
heat kernel theory on manifolds with boundaries for a wide class of
generalized Neumann boundary conditions. In particular, we easily
recover several lowest order surface terms in the case of Robin and
oblique boundary onditions. We briefly discuss multi-loop
applications of the suggested Dirichlet reduction and the prospects
of constructing the universal background field method for systems
with branes/boundaries, analogous to the Schwinger-DeWitt technique.
\end{abstract}

\section{Introduction}
Prospective goal of the present paper is to develop background field
method \cite{DeWitt,DeWitt1} for brane models in gravity/string
theory and cosmology. Current status of brane concept essentially
relies on the analysis of quantum properties beyond the tree-level
approximation. This is especially important in the problem of low
strong coupling scale in brane induced gravity models which
incorporate an infinite sequence of strongly coupled higher
dimensional operators. Their calculation is most efficient in the
language of quantum brane effective action. This, in its turn,
requires application of the covariant method of background field
\cite{DeWitt,DeWitt1} in which the propagators of the theory within
certain approximation are calculable in the external (mean) fields
of a generic form.

The peculiarity of brane models is that their bulk propagators,
rather than being defined in infinite spacetime with simple falloff
conditions, are subject to nontrivial boundary conditions on branes.
Since the fields are subject to dynamical quantum fluctuations on
timelike branes, these boundary conditions belong to the class of
generalized Neumann boundary conditions involving on branes the
values of fields together with their normal and tangential
derivatives. Finding such propagators (usually based on the method
of images) is a very hard task, especially for fields with spins,
when their Green's functions have numerous spin-tensor indices.

On the contrary, Green's functions with Dirichlet boundary
conditions are much easier to obtain -- the method of images gives
them as relatively simple linear combinations of known propagators
in spacetime without boundaries (defined by mirror image
continuation across the original boundary \cite{McKean-Singer}). It
turns out that Feynman diagrammatic technique based on Neumann type
Green's function can be systematically reduced to that of the
Dirichlet type, and the goal of the present paper is to develop the
needed technique. This technique is based on the duality between the
Dirichlet and Neumann boundary value problems, recently discovered
in \cite{duality} at the tree level. Here it will be extended to the
one-loop and multi-loop levels.

The action of a (free field) brane model generally contains the bulk
and the brane parts
    \begin{eqnarray}
    S[\,\phi\,]=\frac12\int\limits_B dX\,\phi(X)\!
    \stackrel{\leftrightarrow}{F}\!(\nabla)\,\phi(X)
    +\int\limits_b
    dx \left(\frac12\,\varphi(x)\,
    \kappa(\partial)\,\varphi(x)+j(x)\,
    \varphi(x)\right),                 \label{1}
    \end{eqnarray}
where the bulk $(d+1)$-dimensional and the brane $d$-dimensional
coordinates are labeled respectively by $X=X^A$ and $x=x^\mu$, and
the boundary values of bulk fields $\phi(X)$ on the brane/boundary
are denoted by $\varphi(x)$
    \begin{eqnarray}
    \phi(X)\,\Big|_{\,b}\equiv\phi\;\Big|=\varphi(x).  \label{2}
    \end{eqnarray}
The kernel of the bulk Lagrangian in given by the second order
differential operator $F(\nabla)$, whose derivatives
$\nabla\equiv\partial_X$ are integrated by parts in such a way that
they form bilinear combinations of first order derivatives acting on
two different fields (this is denoted by
$\stackrel{\leftrightarrow}{F}\!(\nabla)$). Integration by parts in
the bulk gives nontrivial surface terms on the brane/boundary. In
particular, this operation results for a {\em symmetric} operator
$F(\nabla)$ in the Wronskian relation for generic test functions
$\phi_{1,\,2}(X)$
    \begin{eqnarray}
    \int\limits_B d^{\,d+1}X
    \left(\,\phi_1\stackrel{\rightarrow}{F}\!(\nabla)\phi_2-
    \phi_1\!\stackrel{\leftarrow}{F}\!(\nabla)\,\phi_2\right)=
    -\int\limits_{\partial B} d^{\,d}x
    \left(\,\phi_1\stackrel{\rightarrow}{W}\!
    \phi_2-
    \phi_1\stackrel{\leftarrow}{W}\!
    \phi_2\right).                          \label{3}
    \end{eqnarray}
This relation can be regarded as a definition of the first order
{\em Wronskian operator} $W=W(\nabla)$ on the boundary/brane
$b=\partial B$ of spacetime bulk domain $B$. Arrows everywhere here
indicate the direction of action of derivatives either on $\phi_1$
or $\phi_2$.

The brane part of the action contains as a kernel some local
differential operator $\kappa(\partial)$, $\partial=\partial_x$.
Integration by parts here is irrelevant for our purposes, because
$b$ is considered to be either closed compact or having trivial
vanishing boundary conditions at infinitely remote boundary
$\partial b$. $j(x)$ plays the role of sources located on the
boundary. The order of the operator $\kappa(\partial)$ in
derivatives depends on the model in question. In the Randall-Sundrum
model \cite{RS}, for example, for certain gauges it is just an
ultralocal multiplication operator generated by the tension term on
the brane. In the Dvali-Gabadadze-Porrati model \cite{DGP} this is a
second order operator induced by the brane Einstein term,
$\kappa(\partial)\sim\Box/\mu$ where $\mu$ is a very low DGP scale
of the order of magnitude of the horizon scale, responsible for the
cosmological acceleration \cite{Deffayet}. In context of Born-Infeld
action in D-brane string theory with vector gauge fields
$\kappa(\partial)$ is a first order operator \cite{open}.

In all these cases the action (\ref{1}) with dynamical (not fixed)
boundary conditions $\varphi(x)$ for bulk fields naturally gives
rise to generalized Neumann boundary conditions of the form
    \begin{eqnarray}
    \left.\Big(\stackrel{\rightarrow}{W}\!(\nabla)
    +\kappa(\partial)\Big)\,\phi_N\,\right|
    =0,                                      \label{4}
    \end{eqnarray}
which involve normal derivatives of $\phi(X)$ contained in
$\stackrel{\rightarrow}{W}\!(\nabla)$ and generically also the
tangential to the boundary derivatives contained in
$\kappa(\partial)$ (and possibly in
$\stackrel{\rightarrow}{W}\!(\nabla)$). These boundary conditions
when imposed on all brane boundaries of the bulk along with
regularity requirements at the bulk infinity uniquely define the
Neumann type bulk propagator of the theory, brane-to-brane
propagator, etc. and, therefore, uniquely specify all orders of
perturbation theory for both tree-level and quantum effective
action.

Main result which we want to advocate here is that all the Neumann
type ingredients of the perturbation theory can be systematically
reduced to those of the Dirichlet boundary conditions
    \begin{eqnarray}
    \phi_D\,\Big|=0.               \label{5}
    \end{eqnarray}
In particular, the tree-level brane effective action, obtained from
(\ref{1}) by integrating out the bulk fields subject to boundary
conditions (\ref{2}) reads as
    \begin{eqnarray}
    &&\mbox{\boldmath$S$}_{\,\rm brane}[\,\varphi\,]=
    \frac12\,\int\limits_b dx\,dy\,\varphi(x)\,
    \mbox{\boldmath$F$}^{\,\rm brane}(x,y)\,
    \varphi(y)
    +\int\limits_b dx j(x)\,\varphi(x),     \label{6}\\
    &&\mbox{\boldmath$F$}^{\,\rm brane}(x,y)=-
    \stackrel{\rightarrow}{W}\!G_{D}\!
    \stackrel{\leftarrow}{W}\!||\,(x,y)
    +\kappa(\partial)\,\delta(x,y)           \label{7}
    \end{eqnarray}
with the brane-to-brane operator $\mbox{\boldmath$F$}^{\,\rm
brane}(x,y)$ expressed in terms of the Dirichlet Green's function
$G_D(X,Y)$ of the operator $F(\nabla)$ in the bulk. This expression
implies that the kernel of the Dirichlet Green's function is being
acted upon both arguments by the Wronskian operators with a
subsequent restriction to the brane. Double vertical bar indicates
that both points of the operator kernel are restricted to the brane
and labeled by corresponding low case letters. That is, if the
embedding of the boundary/brane in the bulk is denoted by $X=e(x)$,
then this explicitly means:
    \begin{eqnarray}
    \stackrel{\rightarrow}{W}\!G_{D}\!
    \stackrel{\leftarrow}{W}\!||\,(x,y)\equiv\;
    \stackrel{\rightarrow}{W}\!(\nabla_X\!)\,G_{D}(X,Y)\!
    \stackrel{\leftarrow}{W}\!(\nabla_Y\!)
    \,\Big|_{\,X=e(x),\,Y=e(y)}.             \label{8}
    \end{eqnarray}

It is obvious that $\mbox{\boldmath$F$}^{\,\rm brane}(x,y)$ is
essentially nonlocal, its local part being presented by the last
term of (\ref{7}) -- the contribution from the brane. The Green's
function $\mbox{\boldmath$G$}_{\,\rm brane}(x,y)$ of the brane
operator,
    \begin{eqnarray}
    \int_b dz\,\mbox{\boldmath$F$}^{\,\rm brane}(x,z)\,
    \mbox{\boldmath$G$}_{\,\rm brane}(z,y)=\delta(x,y),       \label{9}
    \end{eqnarray}
is the brane-to-brane propagator of the bulk theory, and with the
conventions of the above type this reads as the following expression
for the brane restriction of the Neumann Green's function $G_N(X,Y)$
of $F(\nabla)$
    \begin{eqnarray}
    G_N ||\,(x,y)=\mbox{\boldmath$G$}_{\,\rm brane}(x,y).  \label{10}
    \end{eqnarray}

The duality relations (\ref{7}) and (\ref{10}) were derived in
\cite{duality} for a simplest case of $\kappa(\partial)=0$. Below we
generalize them to the case of a nontrivial $\kappa(\partial)$ and,
moreover, extend them beyond the tree level. In particular, in the
one-loop approximation we show that the functional determinant of
the bulk operator $F(\nabla)$ subject to the generalized Neumann
boundary conditions (\ref{4}) factorizes into the product of the
Dirichlet type determinant of $F(\nabla)$ and the functional
determinant of the brane-to-brane operator
$\mbox{\boldmath$F$}^{\,\rm brane}$ of the {\em boundary
($d$-dimensional)} theory -- the fact briefly reported in
\cite{gospel}. This implies the following additive property for the
one-loop effective action
    \begin{eqnarray}
    \mbox{\boldmath$\varGamma$}_{\rm 1-loop}\equiv\frac12\,
    {\rm Tr}_N\,\ln F=\frac12\,{\rm Tr}_D\,\ln F
    +\frac12\;{\rm tr}
    \ln \mbox{\boldmath$F$}^{\,\rm brane},    \label{11}
    \end{eqnarray}
where ${\rm Tr}_{D,\,N}$ denotes functional traces of the bulk
theory subject to Dirichlet and Neumann boundary conditions, while
${\rm tr}$ is a functional trace in the boundary $d$-dimensional
theory.

Certainly, beyond tree level the effective action contains
ultraviolet divergences, so that this one-loop Dirichlet-Neumann
reduction property (\ref{11}) should be understood within certain
regularization. Below we will use the dimensional regularization
with the dimension $d$ continued to the complex plane playing the
role of regularization parameter. In principle, other
regularizations are possible, and their admissible types will be
discussed in Sect. 2 below.

As a byproduct of (\ref{11}) we suggest a new technique for surface
terms of the local heat kernel expansion in spacetimes with
boundaries. Heat kernel gives a proper time representation for the
functional determinant of (pseudo)differential operators and,
therefore, serves as a basic tool for the calculation of effective
action in background field formalism
\cite{DeWitt,DeWitt1,PhysRep,CPTI,Vassilevich}. In the presence of
boundaries its local expansion is modified by additional surface
terms which amend easily calculable bulk terms well known in physics
context as Schwinger-DeWitt coefficients. Calculation of these
surface terms \cite{McKean-Singer} presents a strong challenge of
both technical and sometimes conceptual (nonperturbative) nature,
especially for the so called oblique boundary conditions
\cite{Osborn-McAvity} which contain derivatives tangential to the
brane and arise, in particular, in Born-Infeld context \cite{open}.
Interestingly, Neumann-Dirichlet duality relations suggest an
alternative method of their calculation, which in view of simplicity
and universality has essential advantages as compared to the
conventional approach of
\cite{Osborn-McAvity,DowkerKirsten,AvramEsp,Vassilevich}.

The paper is organized as follows. In Sect. 2 we derive the
algorithms (\ref{7}), (\ref{10})-(\ref{11}). In Sect. 3 we
demonstrate them on several simple examples. In Sect. 4 we calculate
several lowest order surface terms of the heat kernel for a wide
class of generalized Neumann boundary conditions including, in
particular, the case of oblique ones. In the concluding Sect.5 we
briefly discuss the extension of the technique to multi-loop orders,
its peculiarities in gauge theories and other problems of the
curvature expansion in spacetimes with boundaries, which will be
fully considered in the forthcoming papers \cite{qeastbg,progress}.
In Appendix A the classical Feynman derivation of the gaussian
functional integral in spacetime with boundaries is briefly
revisited, and Appendix B is devoted to the derivation of the heat
kernel for a special case of the brane-to-brane operator.

\section{Neumann vs Dirichlet problems}
We begin this section with specifying in more detail the structure
of the second order bulk operator $F(\nabla)$. As a kernel of the
bulk action in (\ref{1}) it should be symmetric and have the
following general form
    \begin{eqnarray}
    F(\nabla)
    =-\partial_A a^{AB}\partial_B
    -b^A\partial_A+\partial_A(b^A)^T+c.  \label{2.1}
    \end{eqnarray}
Its coefficients are some general coordinate dependent matrices
acting in the vector space of $\phi(X)$ labeled by some spin-tensor
indices which we do not specify here. With respect to these indices
the coefficients $a^{AB}$ and $c$ are symmetric $(a^{AB})^T=a^{AB}$,
$c^T=c$.

The Lagrangian of the bulk part of (\ref{1}) for this operator,
containing the first-order derivatives, is of the form
    \begin{eqnarray}
    \frac12\,\phi
    \stackrel{\leftrightarrow}{F}(\nabla)\,\phi
    =\frac12\,\partial_A\phi\,a^{AB}\partial_B\phi
    -\phi\,b^A\partial_A\phi
    +\frac12\,\phi\,c\,\phi.                     \label{2.2}
    \end{eqnarray}
With one integration by parts, this Lagrangian differs by the total
derivative term from the expression in which the operator
$F(\nabla)$ acts entirely to the right. For two different test
functions $\phi_{1,\,2}$ this reads as
    \begin{eqnarray}
    \phi_1
    \stackrel{\leftrightarrow}{F}\phi_2=
    \phi_1(\stackrel{\rightarrow}{F}\!\phi_2)
    +\partial_A\!\left(\phi_1
    \stackrel{\rightarrow}{W}\!{\vphantom 1}^A
    \phi_2\right)                                 \label{2.3}
    \end{eqnarray}
in terms of the {\em local} Wronskian operator
    \begin{eqnarray}
    \stackrel{\rightarrow}{W}\!{\vphantom 1}^A(\nabla)=
    a^{AB}\partial_B+b^A                              \label{2.4}
    \end{eqnarray}
and can also be rewritten as a {\em local} Wronskian relation
    \begin{eqnarray}
    \phi_1\stackrel{\rightarrow}{F}\!(\nabla)\phi_2-
    \phi_1\!\stackrel{\leftarrow}{F}\!(\nabla)\,\phi_2=
    -\partial_A\!\left(\,
    \phi_1\stackrel{\rightarrow}{W}\!{\vphantom 1}^A
    \phi_2-
    \phi_1\stackrel{\leftarrow}{W}\!{\vphantom 1}^A
    \phi_2\right).                              \label{2.5}
    \end{eqnarray}
Integrating (\ref{2.3}) over the bulk we have the equation
    \begin{eqnarray}
    &&\int\limits_B d^{\,d+1}X\,\phi_1\!
    \stackrel{\leftrightarrow}{F}\!\phi_2=
    \int\limits_B d^{\,d+1}X\,
    \phi_1(\stackrel{\rightarrow}{F}\!\phi_2)
    +\int\limits_b d^{\,d}x\,\phi_1\!
    \stackrel{\rightarrow}{W}\!
    \phi_2\,\Big|\,.                \label{2.6}
    \end{eqnarray}
It determines the {\em boundary/brane} Wronskian operator
$\stackrel{\rightarrow}{W}$ which is given by the normal projection
of the local operator (\ref{2.4}) at the boundary (up to the measure
factor involving the ratio of determinants of the bulk metric
$G_{AB}$ and induced on the brane metric $g_{\mu\nu}$),
$\stackrel{\rightarrow}{W}=({\sqrt g}/{\sqrt
G})\stackrel{\rightarrow}{W}\!{\vphantom 1}^\perp$. Similar
integration of (\ref{2.5}) yields equation (\ref{3}) of
Introduction\footnote{Note that the Wronskian relation (\ref{3})
specifies $W(\nabla)$ only up to arbitrary symmetric operator acting
the boundary (like $\kappa(\partial)$), while Eq.(\ref{2.6}) fixes
it uniquely.}.

Now consider the functional integral in the brane model with the
action (\ref{1})
    \begin{eqnarray}
    Z=\int D\phi\,\exp{(-S[\phi])},  \label{2.7}
    \end{eqnarray}
where the integration runs over the bulk fields $\phi(X)$ and also
over its boundary values $\varphi(x)$, (\ref{2}), on the timelike
branes. Integration over the latter follows from the dynamical
nature of $\varphi(x)$ which are subject to independent quantum
fluctuations. This gaussian path integral equals
    \begin{eqnarray}
    Z=\left(\,{\rm Det}\,G_N\right)^{1/2}
    \exp(-S[\,\phi_N\,]),                   \label{2.8}
    \end{eqnarray}
where $\phi_N$ is a stationary point of the action (\ref{1})
satisfying the following problem with the inhomogeneous Neumann
boundary conditions
    \begin{eqnarray}
    &&F(\nabla)\,\phi_N(X)=0,  \nonumber\\
    &&\left.(\,\stackrel{\rightarrow}{W}
    +\,\kappa\,)\,\phi_N\,\right|+j(x)=0,  \label{2.9}
    \end{eqnarray}
and $G_N$ is the Neumann Green's function of the bulk operator --
the solution of the following problem
    \begin{eqnarray}
    &&F(\nabla)\,G_N(X,Y)=\delta(X,Y),\nonumber\\
    &&(\,\stackrel{\rightarrow}{W}
    +\kappa\,)\,G_N(X,Y)\,\Big|_{\,b}=0.               \label{2.10}
    \end{eqnarray}

For completeness in Appendix A we present the derivation of the
gaussian integral (\ref{2.8}) by Feynman's method \cite{Feynman}
which clearly shows how the boundary conditions enter the
calculation of the preexponential part of this algorithm. This
derivation, in particular, gives the variational definition of the
corresponding functional determinant which goes far beyond its
matrix (finite-dimensional) analogue. It is important that the
boundary value problem (\ref{2.9}) naturally follows from the action
(\ref{1}) and Wronskian relations for $F(\nabla)$, because the
variation of the action is given by the sum of bulk and brane terms
    \begin{eqnarray}
    \delta S[\,\phi\,]=\int_B dX\,\delta\phi\,
    (\stackrel{\rightarrow}{F}\!\phi)+
    \int_b dx\,\delta\varphi\,
    (\stackrel{\rightarrow}{W}\!\phi+\kappa\phi+j)\,\Big|  \label{2.11}
    \end{eqnarray}
which separately should vanish (remember that the action should be
stationary also with respect to arbitrary variations of the boundary
fields $\delta\varphi$).

The solution of (\ref{2.9}) has the following form in terms of the
Neumann Green's function
    \begin{eqnarray}
    &&\phi_N(X)=-\int_b dy\,G_N(X,y)\,j(y)\equiv
    -G_N|\,j,                          \label{2.12}\\
    &&G_N(X,y)\equiv G_N(X,Y)\,\Big|_{\,Y=e(y)}, \nonumber
    \end{eqnarray}
and the stationary action as a functional of the boundary source
$j(x)$ equals
    \begin{eqnarray}
    &&S[\,\phi_N\,]=\frac12\int_B dX\,\phi\,
    (\stackrel{\rightarrow}{F}\phi)
    +\int_b dx \left(\,\frac12\,\phi\,
    (\stackrel{\rightarrow}{W}
    +\kappa)\,\phi
    +j\,\varphi\right),\nonumber\\
    &&\nonumber\\
    &&\qquad\qquad\qquad\qquad
    =-\frac12 \int_b dx\,dy\,
    j(x)\,G_N(x,y)\,j(y)\equiv
    -\frac12 \left.\left.
    j\,G_N\right|\right|\,j\\
    &&G_N(x,y)\equiv G_N(X,Y)\,
    \Big|_{\,X=e(x),\,Y=e(y)}\equiv
    G_N||\, .                                        \label{2.13}
    \end{eqnarray}
Here to simplify the formalism we used condensed notations by
omitting the sign of integration over {\em boundary/brane
coordinates}\,\footnote{We will never use this rule for bulk
integration which will always be explicitly indicated together with
the corresponding integration measure. It is useful to apply this
rule for integral operations on the brane, though, because these
operations never lead to surface terms and in our context have
properties of formal matrix contraction and multiplication. In the
case of the one-dimensional bulk this rule applies literally, and it
can be extended to higher dimensions without any risk of
confusion.}. Thus finally we have
    \begin{eqnarray}
    Z=\left(\,{\rm Det}\,G_N\right)^{1/2}
    \exp\Big(\frac12 \left.\left.
    j\,G_N\right|\right|\,j\Big).                 \label{2.14}
    \end{eqnarray}

Alternatively one can calculate the same integral by splitting the
integration procedure into two steps -- first integrating over bulk
fields with fixed boundary values followed by the integration over
the latter
    \begin{eqnarray}
    \int D\phi\,(...)=\int d\varphi\,
    \int\limits_{\phi|\,=\,\varphi} D\phi\,(...)\,. \label{2.15}
    \end{eqnarray}
This allows one to rewrite the same result in the form
    \begin{eqnarray}
    &&Z=\int d\varphi\,Z(\,\varphi\,),            \label{2.16}\\
    &&Z(\,\varphi\,)=
    \int\limits_{\phi|\,=\,\varphi}
    D\phi\,\exp{(-S[\,\phi\,])},              \label{Z}
    \end{eqnarray}
where the inner integral over bulk fields in view of gaussianity is
again given by the contribution of a saddle point $\phi_D$
    \begin{eqnarray}
    Z(\,\varphi\,)=
    \left(\,{\rm Det}\,G_{D}\right)^{1/2}
    \exp(-S[\,\phi_{D}]).            \label{Z1}
    \end{eqnarray}
This saddle point configuration satisfies the problem with the
inhomogeneous Dirichlet boundary conditions
    \begin{eqnarray}
    &&F(\nabla)\,\phi_D(X)=0, \nonumber \\
    &&\phi_D\,|=\varphi(x),           \label{2.17}
    \end{eqnarray}
and the preexponential factor of (\ref{Z1}) is given by the
functional determinant of the Dirichlet Green's function subject to
    \begin{eqnarray}
    &&F(\nabla)\,G_D(X,Y)=\delta(X,Y),\nonumber\\
    &&G_D(X,Y)\,\Big|_{\,X}=0.            \label{2.18}
    \end{eqnarray}

In terms of this Green's function and using condensed notations we
have
    \begin{eqnarray}
    &&\phi_D(X)=-\int_b dy\;G_D(X,Y)\!
    \stackrel{\leftarrow}{W}\Big|_{\,Y=e(y)}\varphi(y)\equiv
    -G_D\!
    \stackrel{\leftarrow}{W}\!|\;\varphi,         \label{2.19}\\
    &&\nonumber\\
    &&S[\,\phi_D\,]=
    \frac12\,\int_b dx\,dy\,\varphi(x) \Big[\,
    -\stackrel{\rightarrow}{W}\!G_{D}\!
    \stackrel{\leftarrow}{W}(x,y)+\kappa(x,y)\,\Big]\,
    \varphi(y)
    +\int_b dx\,j(x)\,\varphi(x)\nonumber\\
    &&\nonumber\\
    &&\qquad\qquad\qquad\qquad\qquad\equiv
    \frac12\,\varphi\Big[\,-
    \stackrel{\rightarrow}{W}\!G_{D}\!
    \stackrel{\leftarrow}{W}\!||+\kappa\,\Big]\,
    \varphi
    +j\,\varphi,                              \label{2.20}
    \end{eqnarray}
where $\stackrel{\rightarrow}{W}\!G_{D}\!
\stackrel{\leftarrow}{W}\!||$ is defined by Eq.(\ref{8}) in
Introduction. Note that the last expression is exactly the
tree-level brane effective action obtained from the original action
(\ref{1}) by integrating out the bulk fields subject to boundary
conditions $\varphi(x)$,
    \begin{eqnarray}
    \mbox{\boldmath$S$}_{\,\rm brane}[\,\varphi\,]
    =S\left[\,\phi_D[\varphi]\,\right].                      \label{2.20a}
    \end{eqnarray}

Substituting (\ref{Z1}) with (\ref{2.20}) into (\ref{2.16}) we again
obtain the gaussian integral which is saturated by the saddle point
$\varphi_0$ of the above brane action (\ref{2.20})
    \begin{eqnarray}
    \varphi_0=-\Big[\,-\stackrel{\rightarrow}{W}\!G_{D}\!
    \stackrel{\leftarrow}{W}\!||+\kappa\,\Big]^{-1}\,j\,, \label{2.21}
    \end{eqnarray}
and the final result reads
    \begin{eqnarray}
    &&Z=\Big(\,{\rm Det}\,G_{D}\Big)^{1/2}
    \Big(\,{\rm det}\,
    [-\stackrel{\rightarrow}{W}\!G_{D}\!
    \stackrel{\leftarrow}{W}\!||+\kappa\,]
    \,\Big)^{-1/2}\nonumber\\
    &&\qquad\qquad\qquad\qquad\qquad\qquad
    \times\exp\left(\,\frac12\,j\,
    [\,-\stackrel{\rightarrow}{W}\!G_{D}\!
    \stackrel{\leftarrow}{W}\!||
    +\kappa\,]^{-1}\,j\right),                     \label{2.22}
    \end{eqnarray}
where ${\rm det}$ denotes the functional determinants in the
$d$-dimensional boundary theory.

Comparison of its tree-level and one-loop (preexponential) parts
with those of (\ref{2.14}) then immediately yields two relations
    \begin{eqnarray}
    &&G_N\,||=[\,-\stackrel{\rightarrow}{W}\!G_{D}\!
    \stackrel{\leftarrow}{W}\!||+\kappa\,]^{-1}\equiv
    \mbox{\boldmath$G$}_{\,\rm brane},       \label{2.23}\\
    &&\Big(\,{\rm Det}\,G_{N}\Big)^{1/2}=
    \Big(\,{\rm Det}\,G_{D}\Big)^{1/2}\, \Big(\,{\rm det}\,
    [-\stackrel{\rightarrow}{W}\!G_{D}\!
    \stackrel{\leftarrow}{W}\!||+\kappa\,]
    \,\Big)^{-1/2}.                                 \label{2.24}
    \end{eqnarray}
They underly the algorithms (\ref{6}) - (\ref{7}),
(\ref{10})-(\ref{11}) advocated in Introduction and give the
possibility in a systematic way to express all the quantities in
brane theory in terms of the objects subject to Dirichlet boundary
conditions.

As mentioned in Introduction, the relation between functional
determinants (\ref{2.24}) should be understood within some
ultraviolet regularization. It should regulate the both bulk
functional integrals (\ref{2.7})-(\ref{2.8}) and
(\ref{Z})-(\ref{Z1}) as well as the boundary functional integral
over $\varphi=\varphi(x)$ (of $d$-dimensional theory) in
(\ref{2.15})-(\ref{2.16}). The simplest procedure which regularizes
all three integrations without violating the basic relation
(\ref{2.15}) consists in the analytic continuation in $d$ to the
domain of convergence of relevant Feynman integrals. Other types of
regularization do not seem to violate (\ref{2.15}) too, however they
can qualitatively change the setting of the boundary value problem
underlying the obtained algorithms. For example, the regularization
by higher derivatives, as well as certain versions of the
Pauli-Villars regularizations, increases the order of differential
operators. This changes the order of the normal derivative in the
boundary conditions (\ref{4}) and even the number of the latter, so
that the Dirichlet-Neumann reduction stops working or has to be
essentially modified. For this reason, in what follows we will use
dimensional regularization as the simplest and most efficient
scheme. As we shall see in Sect. 4, it correctly generates the
ultraviolet finite heat kernel underlying the calculation of
functional determinants by the Schwinger proper time method and,
thus, confirms the validity of the chosen regularization technique.

\section{Simple examples}
\subsection{One-dimensional problem}
The simplest case of the Dirichlet-Neumann duality relation can be
demonstrated on the example of the one-dimensional Sturm-Liouville
problem on the segment of finite length $y_+ -y_-=l$
    \begin{eqnarray}
     F(\nabla)=m^2-\frac{\partial^2}
     {\partial y^2},\,\,\,y_-\leq y\leq y_+.   \label{3.1}
    \end{eqnarray}
Its Wronskian operator on the two boundaries of this segment is
given by
    \begin{eqnarray}
    \stackrel{\rightarrow}{W}\!\Big|_{\,\pm}=\pm\,\partial_y,  \label{3.2}
    \end{eqnarray}
and the Dirichlet and Neumann Green's functions are correspondingly
    \begin{eqnarray}
     &&G_D(y,y')=
     -\frac{\sinh{m(y'\!-\!y_+)}\,\sinh{m(y\!-\!y_-)}}
     {m\:\sinh{ml}}\;\theta(y'\!-\!y)
     +(y\leftrightarrow y'),                           \label{3.3}\\
     &&G_N(y,y')=
     \frac{\cosh{m(y'\!-\!y_+)}\,\cosh{m(y\!-\!y_-)}}
     {m\:\sinh{ml}}\;\theta(y'\!-\!y)
     +(y\leftrightarrow y')                            \label{3.4}.
    \end{eqnarray}

Since the boundary of the one-dimensional bulk consists of two
points $y_\pm$, the full brane operator (\ref{7}) has the form of
the two-dimensional matrix with the elements corresponding to the
$\pm$ entries on the two zero-dimensional "branes" (for simplicity
we take the case of $\kappa=0$)
   \begin{eqnarray}
     \mbox{\boldmath$F$}^{\,\rm brane}=
     \frac{m}{\sinh{ml}}
     \left(\,\,
     \begin{array}{cc}
      \cosh{ml} & -1 \\
      &\\
      \!-1 &\,\cosh{ml}
     \end{array}\,\,\right).          \label{3.5}
    \end{eqnarray}
On the other hand, the restriction of the Neumann Green's function
(\ref{3.4}) to the boundary is given by
   \begin{eqnarray}
     G_N\,||=
     \frac{1}{m\,\sinh{ml}}
     \left(\,\,
     \begin{array}{cc}
      \cosh{ml} &  \,1 \\
      &\\
      \!1 & \cosh{ml}\!
     \end{array}\,\,\right).          \label{3.6}
    \end{eqnarray}
This is a matter of a simple verification to check that these two
matrices are inverse to one another which is just the relation
(\ref{9}).

To check the one-loop duality relation one can write the Dirichlet
and Neumann functional determinants of $F(\nabla)$ as products of
eigenvalues of the corresponding spectra. Interestingly, for this
simple problem the Dirichlet spectrum
    \begin{eqnarray}
    &&F(\nabla)\,\phi_k^D=\lambda_k^D\,\phi_k^D,\,\,\,
    \phi_k^D(y_\pm)=0,\\
    &&\lambda_k^D=\frac{\pi^2\,k^2}{l^2}+m^2,\,\,\,k=1,2,3,...
    \end{eqnarray}
coincides with the Neumann spectrum
    \begin{eqnarray}
    &&F(\nabla)\,\phi_k^N=\lambda_k^N\,\phi_k^N,\,\,\,
    \partial_y\phi_k^N(y_\pm)=0,\\
    &&\lambda_k^N=\frac{\pi^2\,k^2}{l^2}+m^2,\,\,\,k=0,1,2,3,...
    \end{eqnarray}
except one eigenmode $k=0$. This constant mode, $\phi_0^N(y)={\rm
const}$, is just absent in the spectrum of the Dirichlet problem.
Therefore
    \begin{equation}
     {\rm Det}_N\,F(\nabla) = {\rm Det}_D\,F(\nabla)\,\lambda_0,
    \end{equation}
and this immediately confirms the relation (\ref{2.24}) in view of
the fact that
    \begin{equation}
     {\rm det}\,\mbox{\boldmath$F$}^{\,\rm brane}=m^2=\lambda_0.
    \end{equation}

\subsection{Half-space with Killing symmetry in extra dimension}
Another example is of field-theoretic nature. It corresponds to
$(d+1)$-dimensional half-space with the $d$-dimensional boundary
plane. Let the operator be given by the $(d+1)$-dimensional
d'Alembertian with mass and let the boundary be located at the
position $y=0$ of the extra-dimensional coordinate $y=X^{d+1}$,
    \begin{eqnarray}
    &&F(\nabla)=m^2-\Box^{(\,d+1)}=m^2-\Box-\partial_y^2,\,\,\,
    \Box\equiv\Box^{(d)},\\
    &&X^A=(x^\mu,\,y),\,\,\,y\geq 0,\\
    &&X|_{\,b}=(x,0)
    \end{eqnarray}
The $d$-dimensional part of the full d'Alembertian can in principle
be curved and nontrivially depending on $x$-coordinates. We only
assume the possibility of separation of variables, so that $y$ is a
Killing direction in the bulk.

For such a setting the Wronskian operator is given by the normal
derivative with respect to the outward-pointing normal and equals
    \begin{eqnarray}
    \stackrel{\rightarrow}{W}=-\partial_y,
    \end{eqnarray}
while the exact Dirichlet and Neumann Green's functions are
    \begin{eqnarray}
    &&G_{D,\,N}(y,y')=\frac1\Delta\,
    \left(e^{-\Delta|\,y-y'|}
    \mp e^{-\Delta(y+y')}\right),  \label{3.30}\\
    &&\Delta\equiv\sqrt{m^2-\Box},   \label{3.31}
  \end{eqnarray}
where the minus and plus signs refer respectively to the Dirichlet
and Neumann cases.

According to (\ref{7}) the brane-to-brane operator equals
    \begin{eqnarray}
    \mbox{\boldmath$F$}^{\,\rm brane}=-
    \stackrel{\rightarrow}{\partial}_y\!G_{D}(y,y')\!
    \stackrel{\leftarrow}{\partial_y}\!|_{\,y=y'=0}=\Delta, \label{3.32}
    \end{eqnarray}
while from (\ref{3.30})
    \begin{eqnarray}
    G_N||=G_N(0,0)=\frac1\Delta,   \label{3.33}
    \end{eqnarray}
which confirms the relations (\ref{7}) and (\ref{10}).

To check (\ref{11}) let us write a variational definition of the
functional determinants for both Dirichlet and Neumann boundary
conditions with respect to general variations of the $d$-dimensional
part of the full operator $\delta F=-\delta\,\Box$. We have
    \begin{eqnarray}
    \delta\,\ln\,{\rm Det}\,G_{D,\,N}=
    -{\rm Tr}\,G_{D,\,N}\,\delta F={\rm
    tr}\,\int_0^\infty dy\,G_{D,\,N}(y,y)\,\delta\,\Box, \label{3.34}
    \end{eqnarray}
where we decomposed the $(d+1)$-dimensional functional trace into
the operation of integrating over $y$ the coincidence limit of the
corresponding $y$-dependent kernel and the $d$-dimensional
functional trace ${\rm tr}$. Then, substituting (\ref{3.30}) we have
    \begin{eqnarray}
    \delta\,\left(\,\ln\,{\rm Det}\,G_{N}-\ln\,{\rm
    Det}\,G_{D}\right)={\rm tr}\,\int_0^\infty dy\,\frac1\Delta\,
    e^{-2\Delta y}\,{\delta\,\Box}=-\delta\,\ln \det \Delta,
    \end{eqnarray}
which, in view of (\ref{3.32}), in the infinitesimal variational
form fully confirms (\ref{11}).

\section{Boundary terms of the local heat kernel expansion}
Neumann-Dirichlet duality can be used for the calculation of the
boundary terms in the local expansion of the heat kernel. In the
absence of boundaries with trivial falloff conditions at infinity,
the heat kernel
    \begin{eqnarray}
    K(\,s\,|\,x,y\,)=e^{-s F(\nabla)}\delta(x,y)   \label{4.1}
    \end{eqnarray}
turns out to be a very efficient tool of the covariant diagrammatic
technique for quantum effective action in curved spacetime and in
external fields of a very generic form
\cite{DeWitt,PhysRep,CPTI,Vassilevich}. Its efficiency is based on
the possibility of expanding this kernel in asymptotic series in
integer powers of $s\to 0$ with the coefficients $\hat{a}_n(X,Y)$
which satisfy simple recurrent equations. These coefficients, often
called in the physics context the Schwinger-DeWitt or HAMIDEW
coefficients, can be explicitly found in the coincidence limit $y=x$
as local invariants built in terms of spacetime curvature of the
bulk metric $G_{AB}$, fibre bundle connection and other background
fields. Thus, they give rise to local low energy expansion of the
effective action in inverse powers of the mass parameter
$m^2\to\infty$, when the inverse propagator of the theory is
supplied with the mass term, $F(\nabla)\to F(\nabla)+m^2$.

In spacetime with boundaries the situation becomes more complicated,
because, similarly to Green's functions, the heat kernel should be
obtained from that of the infinite spacetime by the method of
images. This leads to the expansion of the functional trace of the
heat kernel in half-integer powers of the proper time parameter
\cite{McKean-Singer,BransonGilkey,BranGilkeyVas,Vassilevich}
    \begin{eqnarray}
    {\rm Tr}_{(\,d+1)}\,e^{-sF(\nabla)-s\,m^2}=
    \frac1{(4\pi s)^{(d+1)/2}}\, e^{-s\,m^2}
    \sum\limits_{n=0}^\infty
    \left(\,s^n\,A_n+s^{n/2}\,B_{n/2}\,\right).   \label{4.2}
    \end{eqnarray}
Together with the volume (bulk) terms of the Schwinger-DeWitt
type\footnote{In this section we denote the functional trace in the
$(d+1)$-dimensional bulk and on the $d$-dimensional boundary
respectively by ${\rm Tr}_{(\,d+1)}$ and ${\rm Tr}_{(\,d)}$, while
the notation ${\rm tr}$ is reserved for the trace over spin-tensor
indices of matrices. The latter are denoted by hats.},
    \begin{eqnarray}
    &&A_n=\int\limits_{\cal M}
    d^{\,d+1}X\, G^{1/2}(X)\;
    {\rm tr}\,\hat{a}_n(X,X),                          \label{4.3}
    \end{eqnarray}
this expansion acquires surface integrals at the boundary $B_{n/2}$
which are built of local invariants incorporating also such local
characteristics of the surface as its extrinsic curvature
$K_{\mu\nu}$ and the curvature of the induced metric $g_{\mu\nu}$,
    \begin{eqnarray}
    &&B_{n/2}=\int\limits_{\partial\cal M} d^{d}x\,
    g^{1/2}(x)\,b_{n/2}(x).                       \label{4.4}
    \end{eqnarray}
Here $G(X)$ and $g(x)$ denote the determinants of the bulk and brane
metrics, so that ${\rm tr}\,\hat{a}_n(X,X)$ and $b_{n/2}(x)$ turn
out to be the bulk and boundary scalars.

In contrast to the bulk Schwinger-DeWitt coefficients
$\hat{a}_n(X,X)$ which are universal and independent of the type of
boundary conditions, the surface terms essentially depend on the
latter, and their calculation is much less universal and often very
cumbersome. For the operator of the form
    \begin{eqnarray}
    F(\nabla)=-\,G^{AB}\,\nabla_A\nabla_B
    -\hat P+\frac16\,R\,\hat 1,                 \label{4.5}
    \end{eqnarray}
with second order covariant derivatives $\nabla_A$ forming a
covariant $(d+1)$-dimensional D'Alembertian in the metric $G_{AB}$
and $\hat P$ denoting some matrix-valued potential term, few lowest
Schwinger-DeWitt coefficients read as
    \begin{eqnarray}
    \hat a_0(X,X)=\hat 1,\,\,
    \hat a_1(X,X)=\hat P,\,\,\ldots\,.              \label{4.6}
    \end{eqnarray}
The corresponding surface terms for the Dirichlet and simple Neumann
boundary conditions
    \begin{eqnarray}
    &&b_0^{D,\,N}(x)=0,                   \label{4.7}\\
    &&b_{1/2}^D(x)=
    -\frac{\sqrt\pi}2\,{\rm tr}\,\hat1,\,\,\,\,
    b_{1/2}^N(x)=
    \frac{\sqrt\pi}2\,{\rm tr}\,\hat1,         \label{4.8}\\
    &&b_1^{D,\,N}(x)=
    \frac13\,K\,{\rm tr}\hat1,           \label{4.9}\\
    && \cdots \nonumber
    \end{eqnarray}
involve the trace of the extrinsic curvature of the boundary
$K=g^{\mu\nu}K_{\mu\nu}$ \cite{McKean-Singer,Vassilevich}.

For the generalized Neumann (Robin) boundary conditions
    \begin{eqnarray}
    (\,\nabla_n-\hat S\,)\,
    \phi(X)\,\Big|_{\,\partial\cal M}=0             \label{4.10}
    \end{eqnarray}
the last of the above coefficients is modified by the matrix-valued
potential term $\hat S$ \cite{Vassilevich}
    \begin{eqnarray}
    b_1^R(x)={\rm tr}\left[\,2\,
    \hat S+\frac13\,K\,\hat1\,\right].         \label{4.11}
    \end{eqnarray}

This modification is even more sophisticated in the case of the
so-called oblique boundary conditions \cite{Osborn-McAvity}, which
include the tangential to the boundary ($d$-dimensional) covariant
derivatives $D_\mu$
    \begin{eqnarray}
    \Big(\,\nabla_n-\hat\varGamma^\mu D_\mu
    -\frac12\,(D_\mu \hat\varGamma^\mu)-\hat
    S\Big)\,\phi(X)\,\Big|_{\,\partial\cal M}=0.   \label{4.12}
    \end{eqnarray}
These derivatives enter the boundary conditions with the {\em
dimensionless} matrix-valued vector coefficients
$\hat\varGamma^\mu$. For the generic case their contribution to
$b_{n/2}$ is not known, but in the case of commuting matrices
    \begin{eqnarray}
    [\,\hat\varGamma^\mu,\hat\varGamma^\nu]=0  \label{4.13}
    \end{eqnarray}
lengthy calculations of \cite{Osborn-McAvity,DowkerKirsten,AvramEsp}
lead to the following expressions for few lowest-order surface
densities (see also \cite{DowkerKirsten,Vassilevich} for higher
order $b_{n/2}$)
    \begin{eqnarray}
    &&b_{1/2}^O(x)=\frac{\sqrt\pi}2\,{\rm
    tr}\,\left[\,\frac2{\sqrt{1
    +\hat\varGamma^2}}-\hat1\,\right],              \label{4.14}\\
    &&b_1^O(x)={\rm tr}\left[\,\frac2{1
    +\hat\varGamma^2}\,
    \hat S+\frac13\,K\,\hat1\right.\nonumber\\
    &&\qquad\qquad\qquad\qquad\left.
    +\Big(\,\frac1{1+\hat\varGamma^2}
    -\frac{{\rm arctanh}
    \sqrt{-\hat\varGamma^2}}{\sqrt{-\hat\varGamma^2}}\,\Big)\,
    \Big(\,K-K^{\mu\nu}\frac{\hat\varGamma_\mu
    \hat\varGamma_\nu}{\hat\varGamma^2}\,\Big)\,
    \right],                                          \label{4.15}
    \end{eqnarray}
where
    \begin{eqnarray}
    \hat\varGamma^2=\hat\varGamma^\mu\hat\varGamma_\mu.  \label{4.16}
    \end{eqnarray}
It is important to note that these matrix functions are
nonpolynomial in $\hat\varGamma^\mu$ because of the dimensionless
nature of these matrices. This means that their contribution to any
given surface coefficient $b_{n/2}$ cannot be obtained by the
perturbation theory in $\hat\varGamma^\mu$, which is the main reason
of difficulties in their derivation.

Let us now give a simple derivation of these coefficients by the
technique of the previous section. To begin with, note that the
Wronskian operator for (\ref{4.5}) is defined by the normal
derivative with respect to outward-pointing normal $n^A$ to the
boundary
    \begin{eqnarray}
    \stackrel{\rightarrow}{W}=\nabla_n,\,\,\,
    \nabla_n=n^A\nabla_A.                      \label{4.17}
    \end{eqnarray}
Introduce the covariant operator $\kappa(D)$ corresponding to the
generalized Neumann boundary condition (\ref{4.12}) and the
$d$-dimensional brane action, from which these boundary condition
can be obtained by the variational procedure of Sect.2. Obviously,
they read
    \begin{eqnarray}
    &&\kappa=\kappa(D)=-\hat\varGamma^\mu D_\mu
    -\frac12\,(D_\mu \hat\varGamma^\mu)-\hat S,\\   \label{4.18}
    &&\nonumber\\
    &&S^{(d)}[\,\varphi\,]=\frac12\,\varphi\,\kappa\,\varphi
    =-\frac12\,\int_b
    dx\,g^{1/2}\,
    \varphi^T (\hat\varGamma^\mu D_\mu
    +\hat S)\,\varphi(x),                    \label{4.19}
    \end{eqnarray}
provided the following symmetry property of matrices
$\hat\varGamma^\mu$ and $\hat S$ hold (which we assume in what
follows)
    \begin{eqnarray}
    \hat\varGamma^{\mu\,T}=
    -\hat\varGamma^\mu,\,\,\,\,
    \hat S^T=\hat S.                        \label{4.20}
    \end{eqnarray}

According to the Shwinger-DeWitt proper-time method
\cite{DeWitt,PhysRep} the functional determinants of massive
operators are given by the proper time integrals of the
corresponding heat kernels. In view of the representation
(\ref{4.2}) this gives the following inverse mass expansions in the
Dirichlet and generalized Neumann cases
    \begin{eqnarray}
    &&{\rm Tr}_{D,\,N}\ln\,\Big[\,F(\nabla)+m^2\,\Big]
    =-{\rm Tr}_{D,\,N}\int_0^\infty
    \frac{ds}s\;e^{-sF(\nabla)-s\,m^2}\nonumber\\
    &&\nonumber\\
    &&\qquad\quad
    =-\Big(\,\frac{m^2}{4\pi}\,\Big)^{\!(d+1)/2}\,\left[\,\,\sum_{n=0}^\infty
    \frac{\Gamma\left(n-\frac{d+1}2\right)}{m^{2n}}\;A_n+
    \,\sum_{n=1}^\infty
    \frac{\Gamma\left(\frac{n-d-1}2\right)}{m^n}\;
    B_{n/2}^{D,\,N}\right].              \label{4.22}
    \end{eqnarray}
The heat kernel trace (\ref{4.2}) is always ultraviolet finite, and
the one-loop divergences originate from the proper-time integration
diverging at $s=0$. In dimensional regularization (with $d$
analytically continued to the complex plane) they arise here as
gamma function poles for first few terms of nonnegative powers in
the mass parameter (in even dimension $d+1$ for bulk
Schwinger-DeWitt coefficients and both even and odd $d+1$ for
surface terms).

The Dirichlet and Neumann expressions (\ref{4.22}) differ only by
the contributions of surface integrals, their total difference, on
the other hand, being defined from the duality relation (\ref{11}).
Subtracting the Dirichlet version of (\ref{4.22}) from the Neumann
one, one therefore obtains
    \begin{eqnarray}
    {\rm Tr}_{(d)}\ln \mbox{\boldmath$F$}^{\,\rm brane}=
    -\frac12\,\Big(\,\frac{m^2}{4\pi}\,
    \Big)^{\!d/2}\,\sum_{n=1}^\infty
    \frac{\Gamma\left(\frac{n-d-1}2\right)}{m^{n-1}}\,\,\int
    d^dx\,g^{1/2}\,\frac{b^{N}_{n/2}
    -b^{D}_{n/2}}{\sqrt\pi},               \label{4.23}
    \end{eqnarray}
where the brane-to-brane operator $\mbox{\boldmath$F$}^{\,\rm
brane}$ is defined by Eq. (\ref{7}) in Introduction for a particular
case of the boundary operator (\ref{4.19}). Thus, the difference of
the boundary terms for Neumann and Dirichlet cases can be
disentangled from the functional determinant of
$\mbox{\boldmath$F$}^{\,\rm brane}$ in relevant orders of the
$1/m$-expansion. In view of the ultraviolet finiteness of the heat
kernel (finiteness of the coefficients $b^{N,D}_{n/2}$) the
divergences of this functional determinant should have the structure
of gamma-function coefficients in the right hand side of
(\ref{4.23}), which serves as a consistency check of the whole
procedure.

Though the operator $\mbox{\boldmath$F$}^{\,\rm brane}$ is itself
not exactly known, all we need is its inverse-mass expansion which
is equivalent to the bulk-brane curvature expansion of the Dirichlet
Green's function $G_D$ of $F(\nabla)$. Moreover, it is defined on
the brane manifold without a boundary (or with trivial regularity
conditions at infinity). All this essentially facilitates the
solution of the problem. As we show in the following examples,
choosing the brane operator (\ref{7}) for $\kappa(D)$ of the form
(\ref{4.18}) we easily calculate the lowest order surface terms, and
this procedure can undoubtedly be extended to all $b_{n/2}$.

\subsection{Simple Neumann boundary conditions}
For simplest Neumann boundary conditions with $\kappa(D)=0$ the
brane-to-brane operator (\ref{7}) was obtained in the leading order
approximation in Sect.3.2, Eq. (\ref{3.32}). It reads
    \begin{eqnarray}
    \mbox{\boldmath$F$}^{\,\rm brane}=
    \Delta\equiv\sqrt{m^2-\Box}+O[\,R,K\,],   \label{4.24}
    \end{eqnarray}
where $O[\,R,K\,]$ denotes corrections due to the bulk curvature and
extrinsic curvature of the boundary. Therefore, the inverse mass
expansion for its determinant is dominated by
    \begin{eqnarray}
    &&{\rm Tr}_{(d)}\ln \mbox{\boldmath$F$}^{\,\rm brane}=
    \frac12\,{\rm Tr}_{(d)}\ln (m^2-\Box)+O[\,R,K\,]\nonumber\\
    &&\qquad\qquad\qquad\qquad\quad=-\frac12\,
    \Big(\,\frac{m^2}{4\pi}\,\Big)^{d/2}\,
    \Gamma(-d/2)\,\int d^{\,d}x\,g^{1/2}\,
    {\rm tr}\,\hat1+O[\,R,K\,]                     \label{4.25}
    \end{eqnarray}
with $O[\,R,K\,]=O[\,m^{d-1}\,]$. From the $n=1$ term of
(\ref{4.23}) it follows then that
    \begin{eqnarray}
    b^{N}_{1/2}-b^{D}_{1/2}=\sqrt\pi\,{\rm tr}\,\hat1,  \label{4.26}
    \end{eqnarray}
which fully agrees with (\ref{4.8}). The dependence of (\ref{4.25})
on dimensionality (yielding the logarithmic divergence for even $d$)
is exactly the same as in the $n=1$ term of (\ref{4.23}) which, as
expected, guarantees the ultraviolet finiteness of the obtained
difference $b^{N}_{1/2}-b^{D}_{1/2}$.

\subsection{Robin boundary conditions}
For Robin boundary conditions (\ref{4.10}) with $\kappa(D)=-\hat S$
    \begin{eqnarray}
    \mbox{\boldmath$F$}^{\,\rm brane}=
    \sqrt{m^2-\Box}-\hat S+O[\,R,K\,].        \label{4.27}
    \end{eqnarray}
The functional determinant of this operator can be obtained by
perturbation theory in the {\em dimensional} quantity $\hat S$
    \begin{eqnarray}
    &&{\rm Tr}_{(d)}\ln \mbox{\boldmath$F$}^{\,\rm brane}=
    \frac12\,{\rm Tr}_{(d)}\ln (m^2-\Box)-{\rm Tr}_{(d)}\frac{\hat
    S}{\sqrt{m^2-\Box}}
    +O[\,R,K,\hat S^2\,].                      \label{4.28}
    \end{eqnarray}
Only the first order in $\hat S$ contributes to $b^{R}_1$ in the
Robin case, and to zeroth order in the curvature this term equals
    \begin{eqnarray}
    &&-{\rm Tr}_{(d)}\frac{\hat
    S}{\sqrt{m^2-\Box}}=-\frac1{\varGamma(1/2)}\,{\rm
    Tr}_{(d)}\int_0^\infty ds\,s^{-1/2}\,
    e^{-s(m^2-\Box)}\,\hat S
    \nonumber\\
    &&\qquad\qquad\qquad\qquad=
    -\int_0^\infty \frac{ds}{\sqrt{\pi s}}\,
    \frac{e^{-sm^2}}{(4\pi s)^{d/2}}\,
    \int d^d x\,g^{1/2}\,{\rm tr}\,\hat S+O[\,R,K\,]\nonumber\\
    &&\qquad\qquad\qquad\qquad=
    -\Big(\,\frac{m^2}{4\pi}\,\Big)^{d/2}\,
    \frac{\Gamma\left(\frac{1-d}2\right)}{m\sqrt\pi}\,\,
    \int d^d x\,g^{1/2}\,
    {\rm tr}\,\hat S+O[\,R,K\,].           \label{4.29}
    \end{eqnarray}
Therefore, from the $n=2$ term of (\ref{4.23}) it follows that
    \begin{eqnarray}
    b^{R}_1-b^{D}_1=2\,{\rm tr}\,\hat S+O[\,K\,],  \label{4.30}
    \end{eqnarray}
which fully agrees with (\ref{4.9}) and (\ref{4.11}).

\subsection{Oblique boundary conditions}
For simplicity consider the case of oblique boundary conditions
(\ref{4.12}) with $\hat S=0$. Then the surface operator (\ref{4.18})
is $\kappa(D)=-\hat\varGamma^\mu D_\mu+O[\,D\hat\varGamma\,]$, and
the brane-to-brane operator reads
    \begin{eqnarray}
    \mbox{\boldmath$F$}^{\,\rm brane}=
    \sqrt{m^2-\Box}-\hat\varGamma^\mu D_\mu
    +O[\,R,K,D\hat\varGamma\,],                      \label{4.31}
    \end{eqnarray}
where together with curvature terms we disregard terms with
covariant derivatives of $\hat\varGamma^\mu$. With the same accuracy
    \begin{eqnarray}
     &&\mathrm{Tr}\ln\mbox{\boldmath$F$}^{\,\rm brane}=
     -\int d^dx\,\;\mathrm{tr}
     \int\limits_0^\infty \frac{d\tau}{\tau}\;
      e^{-\tau\sqrt{m^2-\Box}
      +\tau\hat\varGamma^\mu D_\mu}\,
      \delta(x,y)\,\Big|_{y=x}\,.                  \label{4.32}
    \end{eqnarray}
Here we denote the proper time parameter by $\tau$ to emphasize that
it has the dimensionality different from $s$ (length rather than
length squared), which in its turn is explained by the
dimensionality of (\ref{4.31}).

In contrast to the Robin case further calculations cannot be
performed by perturbations in powers of the term $\hat\varGamma^\mu
D_\mu$, because the vector coefficient $\hat\varGamma^\mu$ is
dimensionless, and the perturbation theory in this term will not
generate asymptotic expansion in inverse mass\footnote{In other
words, the leading symbol of the pseudodifferential operator
(\ref{4.31}) -- the highest (first) order in derivatives part of
$\mbox{\boldmath$F$}^{\,\rm brane}$ -- is given by
$\sqrt{-\partial^2}-\hat\varGamma^\mu \partial_\mu$, and it should
be treated as a whole without breaking into pieces.}. Instead, with
the same accuracy of zeroth order in the curvature one can
disentangle this term in the exponential as a matrix-valued shift
operator
    \begin{eqnarray}
    &&\exp\left[\,-\tau\sqrt{m^2-\Box}
    +\tau\hat\varGamma^\mu\nabla_\mu\,
    \right]\delta(x,y)\,\Big|_{\,y=x}
    \nonumber\\
    &&\qquad\qquad\qquad\qquad\qquad=
    e^{\tau\hat\varGamma^\mu\nabla_\mu}\;
    K(\,\tau\,|\,x,y)\,\Big|_{\,y=x}=
    K(\,\tau\,|\,x+\tau\hat\varGamma,x)    \label{4.33}
    \end{eqnarray}
acting on the heat kernel of the operator $\sqrt{m^2-\Box}$,
    \begin{equation}
    K(\,\tau\,|\,x,y)=
    \exp\left[\,-\tau\,\sqrt{m^2-\Box}\;
    \right] \delta(x,y).                        \label{4.34}
    \end{equation}
Note that in view of the commutativity assumption (\ref{4.13}) this
nontrivial matrix-valued function is uniquely defined without any
matrix ordering prescription\footnote{Apparently, the commutativity
assumption can be removed. For non-commuting matrices equation
(\ref{4.33}) will still be valid under the symmetric matrix ordering
prescription in the right hand side. This symmetrization follows
from the symmetry of the Taylor series coefficients.}.

As shown in Appendix B, in flat $(d+1)$-dimensional space without
boundaries
    \begin{equation}
    K(\,\tau\,|\,x,y)
    =2\tau \Big(\,\frac{m^2}{2\pi Z}\,
    \Big)^{\!(\,d+1)/2} K_{(\,d+1)/2}(Z),  \label{4.35}
    \end{equation}
where
    \begin{equation}
    Z\equiv Z(\,\tau\,|\,x,y)=
    m\sqrt{|\,x-y|^{\,2}+\tau^{\,2}}          \label{4.36}
    \end{equation}
and $K_\nu(Z)$ is the modified Bessel (MacDonald) function. Because
of
    \begin{eqnarray}
    Z(\,\tau\,|\,x+\tau\hat\varGamma,x)=
    m\,\tau\,\sqrt{1+\hat\varGamma^2}
    \end{eqnarray}
the $\tau$-integration in (\ref{4.32}) gives
    \begin{eqnarray}
     &&\mathrm{Tr}\ln\,
     \mbox{\boldmath$F$}^{\,\rm brane}=
     -\int d^d x\,g^{1/2}\;
     \mathrm{tr}\int\limits_0^\infty
     \frac{d\tau}{\tau}\;
     K(\,\tau\,|\,x+\tau\hat\varGamma,x)\nonumber\\
     &&\qquad\quad
     =-\Big(\,\frac{m^2}{2\pi}\,\Big)^{\!(\,d+1)/2}\frac2{m}\,
     \int d^d x\,g^{1/2}\;\mathrm{tr}\,
    \frac1{\sqrt{1+\hat\varGamma^2}}\,
    \int_0^\infty dZ\,\frac{K_{(\,d+1)/2}(Z)}{Z^{\!(\,d+1)/2}}\,,
    \end{eqnarray}
whence in view of the integral ( Eq. 6.561.16 of
\cite{Gradshtein-Ryzhik})
    \begin{eqnarray}
    \int\limits_0^\infty dx \;x^{-\nu}
    K_\nu(x)=2^{-\nu-1}\sqrt{\pi}\;
    \Gamma\left(\frac{-2\nu+1}{2}\right)
    \end{eqnarray}
we finally have
    \begin{eqnarray}
    {\rm Tr}_{(d)}\ln \mbox{\boldmath$F$}^{\,\rm brane}
    =-\frac12\,\Big(\,\frac{m^2}{4\pi}\Big)^{\!d/2}\,
    \Gamma\Big(\!-\frac d2\Big)\!\int d^d x\,g^{1/2}\,\mathrm{tr}
    \frac1{\sqrt{1+\hat\varGamma^2}}+O[\,m^{d-1}\,].
    \end{eqnarray}
Therefore, from the $n=1$ term of (\ref{4.23}) it follows that
    \begin{eqnarray}
    b^{O}_{1/2}-b^{D}_{1/2}=\sqrt\pi\;{\rm tr}\,
    \frac1{\sqrt{1+\hat\varGamma^2}}
    \end{eqnarray}
which fully agrees with (\ref{4.14}) for oblique boundary conditions
with $\hat S=0$. Similarly, one can check the $\hat S$-dependent
term of (\ref{4.15}) for the case of nonvanishing $\hat S$.

The approximation of zero bulk and brane curvatures in all the above
examples can be used as a starting point of the perturbation theory
in $O[\,R,K\,]$ (and in other dimensional background field
quantities like $D_\mu\hat\varGamma^\mu$ and $\hat P$). Then the
Neumann-Dirichlet duality method of the above type will give
extrinsic curvature terms of (\ref{4.15}) and all higher order
surface terms in the heat kernel expansion (\ref{4.2}).

\section{Conclusions}
Thus, the algorithms (\ref{7}), (\ref{10})-(\ref{11}) allow one to
reduce calculations of brane effective action to those of the
Dirichlet boundary conditions. This reduction technique can
obviously be extended to multi-loop orders by applying the same
trick of splitting the functional integration into two steps, as in
(\ref{2.15}) - (\ref{Z}), also in the nonlinear case. The resulting
Feynman diagrammatic technique from combinatorical viewpoint is not
so simple as in (\ref{11}), but still manageable. In addition to the
bulk Dirichlet type propagator $G_D$ it has the brane-to-brane
propagator $\mbox{\boldmath$G$}_{\,\rm brane}$  -- the Green's
function of $\mbox{\boldmath$F$}^{\,\rm brane}$. The modification of
Feynman diagrams, therefore, consists in the insertions into the
bulk diagrams of the Dirichlet type the lines connecting bulk
vertices to the brane by the brane-to-boundary propagators
    \begin{eqnarray}
    \frac{\delta\phi_D(X)}{\delta\varphi(y)}=-G_D(X,Y)
    \stackrel{\leftarrow}{W}\Big|_{\,Y=e(y)}         \label{5.1}
    \end{eqnarray}
(cf. Eq.(\ref{2.19})) and also developing the $d$-dimensional
diagrammatic technique on the brane with the propagator
$\mbox{\boldmath$G$}_{\,\rm brane}$. This technique will be
considered in more detail in \cite{qeastbg}. In gravitational brane
models the question of gauge invariance (especially with respect to
general coordinate transformations) becomes very important, while at
present even the details of Faddeev-Popov gauge fixing procedure in
spacetimes with boundaries are not clearly studied \cite{gospel}.
Therefore, gauge properties of Neumann-Dirichlet duality  will be a
major aspect of \cite{qeastbg}, where gravitational Ward identities
in brane models will be established (they are briefly reported in
\cite{gospel}).

As a byproduct, the Neumann-Dirichlet reduction technique suggests
also a new method of calculating surface terms of the heat kernel
expansion. Given the Dirichlet type terms, those of the generalized
Neumann case can be obtained from (\ref{4.23}). This allows one to
circumvent the limitations of the conventional method for these
terms. In particular, we were able to recover correct expressions
for few lowest surface contributions to the Schwinger-DeWitt
expansion and, in case of oblique boundary conditions, perhaps even
extend their validity beyond the commutative case (\ref{4.13}) (see
footnote after Eq.(\ref{4.34}) on the symmetric matrix-ordering
prescription).

Of course, the success of Neumann-Dirichlet reduction programme
depends on our ability to find the Dirichlet Green's function $G_D$
and the corresponding brane-to-brane operator
$\mbox{\boldmath$F$}^{\,\rm brane}$. The latter is a nonlocal
pseudodifferential operator, so the problem of efficiently handling
its nonlocality arises. For a wide class of problems
$\mbox{\boldmath$F$}^{\,\rm brane}$ was found in the zero curvature
approximation as
    \begin{eqnarray}
    \mbox{\boldmath$F$}^{\,\rm brane}(D)=
    \sqrt{m^2-\Box}+\kappa(D)+O[\,R,K\,].
    \end{eqnarray}
Despite nonlocality, local expansion of its functional determinant
is still manageable. For ultralocal $\kappa(D)=-\hat S$, as in Robin
case, it is easily available by perturbations. It is more
complicated in the case of oblique boundary conditions, when the
leading symbol of $\mbox{\boldmath$F$}^{\,\rm brane}(D)$  gives rise
to the propagation off light cone on the brane (in the physical
theory with the Lorentzian signature) -- the phenomenon called {\em
generalized causality} in \cite{PhysRep}\footnote{Related phenomenon
of loss of strong ellipticity, which is the unboundedness of the
operator spectrum from below in the Euclidean theory
\cite{ellipticity} or the presence of ghost modes in the Lorentzian
case, takes place when $1+\hat\varGamma^2$ acquires zero or negative
eigenvalues leading to singularities in the algorithms
(\ref{4.14})-(\ref{4.15})}. Finally, in brane induced gravity models
with the operator $\kappa(D)\sim\Box/\mu$ generated by the brane
Einstein term ($\mu$ is the DGP scale \cite{DGP,Deffayet}), the
calculational technique still has to be worked out \cite{progress}.

Final comment of this paper concerns the (bulk and boundary)
curvature expansion of the Dirichlet Green's function. Classical
method of images for this expansion is known \cite{McKean-Singer},
and this method for the Dirichlet case is much simpler than for the
generalized Neumann boundary conditions. Still further efforts are
necessary to convert it into a regular systematic calculational
scheme comparable in its universality to the Schwinger-DeWitt
technique in spacetime without branes/boundaries
\cite{PhysRep,CPTI}. The progress in this direction will be reported
in forthcoming papers \cite{qeastbg,progress}, and here it remains
to express a hope that at least the general shape of
background-field method for quantum effective action in brane theory
becomes visible.

\appendix
\renewcommand{\thesection}{\Alph{section}.}
\renewcommand{\theequation}{\Alph{section}.\arabic{equation}}

\section{Gaussian path integral in spacetime with boundaries}
Feynman's calculation \cite{Feynman} of the gaussian functional
integral (\ref{2.8}) is based on the integral
    \begin{eqnarray}
    Z[\,F,J\,]=\int D\phi\,\exp{(-S[\,\phi,J\,])}.  \label{A.1}
    \end{eqnarray}
Here instead of the source $j(x)$ at the boundary, as in (\ref{1}),
the action has the source $J(X)$ to the integration field $\phi(x)$
in the bulk
    \begin{eqnarray}
    &&S[\,\phi,J\,]=\int_B dX\,\left(\frac12\,\phi(X)\!
    \stackrel{\leftrightarrow}{F}\!(\nabla)\,
    \phi(X)+J(X)\,\phi(X)\right)\nonumber\\
    &&\qquad\qquad\qquad\qquad\qquad\qquad\qquad\quad
    +\frac12\,\int_{\partial B}
    dx \,\varphi(x)\,
    \kappa(\partial)\,\varphi(x).                \label{A.2}
    \end{eqnarray}
To find the dependence of (\ref{A.1}) on $J(X)$ consider the
stationary point of this action with respect to arbitrary variations
of $\phi(X)$ both in the bulk and on the boundary. Similarly to
(\ref{2.9}) this field satisfies the generalized Neumann boundary
value problem
    \begin{eqnarray}
    &&F(\nabla)\,\phi_N(X)+J(X)=0,  \nonumber\\
    &&\left.(\stackrel{\rightarrow}{W}
    +\kappa\,)\,\phi_N\,\right|=0.                 \label{A.3}
    \end{eqnarray}

Now make the shift of the integration variable in (\ref{A.1}) by
$\phi_N$
    \begin{eqnarray}
    \phi=\phi_N+\Delta.
    \end{eqnarray}
Under this replacement the action decomposes in the part
$S[\,\Delta,0\,]$ quadratic in $\Delta$ and the part independent of
$\Delta$. Linear in $\Delta$ term is absent (both in the bulk and on
the boundary) in view of the stationarity of the action at $\phi_N$,
so that
    \begin{eqnarray}
    &&S[\,\phi,J\,]=S[\,\Delta,0\,]+S[\,\phi_N,J\,],   \label{A.4}\\
    &&S[\,\phi_N,J\,]=\frac12\,
    \int_B dX\,dY\,J(X)\,G_N(X,Y)\,J(Y).        \label{A.4a}
    \end{eqnarray}
Therefore
    \begin{eqnarray}
    Z[\,F,J\,]=Z[\,F,0\,]\,\exp{(-S[\,\phi_N,J\,])},  \label{A.5}
    \end{eqnarray}
which justifies the exponential (tree-level) part of (\ref{2.8}).

To find the prefactor, consider the variation of the integral
(\ref{A.1}) at $J=0$ with respect to the operator $F(\nabla)$ and
make the following set of obvious identical transformations using
the above equations (\ref{A.4a}) and (\ref{A.5})
    \begin{eqnarray}
    &&\delta_F Z[\,F,0\,]=-\int D\phi\,
    \left(\frac12\,\int_B dX\,\phi(X)\!
    \stackrel{\leftrightarrow}{\delta F}\!(\nabla)\,
    \phi(X)\!\right)
    \exp{(-S[\,\phi,0\,])}\nonumber\\
    &&\nonumber\\
    &&\qquad\qquad\quad=-\left.\int D\phi\,
    \left(\frac12\,\int_B dX\,\frac\delta{\delta J(X)}\!
    \stackrel{\leftrightarrow}{\delta F}\!(\nabla)\,
    \frac\delta{\delta J(X)}
    \,\right)\,Z[\,F,J\,]\,\right|_{\,J=0}\nonumber\\
    &&\nonumber\\
    &&\qquad\qquad\quad=
    -\frac12\,\int_B dX\,
    \stackrel{\leftrightarrow}{\delta F}\!(\nabla)\,
    G_N(X,Y)\,\Big|_{\,Y=X}\,Z[\,F,0\,].              \label{A.6}
    \end{eqnarray}
Here $\stackrel{\leftrightarrow}{\delta F}\!(\nabla)$ means
arbitrary variations of the coefficients of the operator, $\delta
a^{AB}(X)$, $\delta b^A(X)$, $\delta c(X)$, and the double arrow
implies symmetric action of two first-order derivatives of
$F(\nabla)$ on both arguments of $G_N(X,Y)$ similar to
Eq.(\ref{2.2})
    \begin{eqnarray}
    &&\int_B dX\,
    \stackrel{\leftrightarrow}{\delta F}\!(\nabla)\,
    G_N(X,Y)\,\Big|_{\,Y=X}
    \nonumber\\
    &&\nonumber\\
    &&\qquad\qquad
    \equiv\int_B dX\,\Big[\,
    \delta a^{AB}(X)\,\partial_A^Y\,\partial_B^X
    -2\,\delta b^A(X)\,\partial_A^X
    +\delta c(X)\,\Big]\,G_N(X,Y)\,\Big|_{\,Y=X}\nonumber\\
    &&\nonumber\\
    &&\qquad\qquad\qquad\qquad\qquad\qquad
    ={\rm Tr}\,\stackrel{\leftrightarrow}{\delta F}\!
    (\nabla)\,G_N.                     \label{A.7}
    \end{eqnarray}

Therefore, from (\ref{A.6}) one gets
    \begin{eqnarray}
    &&\delta_F \ln\,Z[\,F,0\,]=
    -\frac12\,{\rm Tr}\,
    \stackrel{\leftrightarrow}{\delta F}\!
    (\nabla)\,G_N =-\frac12\,\delta\ln{\rm Det}_N F\equiv
    \frac12\,\delta\ln{\rm Det}\,G_N\, .           \label{A.8}
    \end{eqnarray}
This expression in the variational form justifies the representation
of the prefactor in (\ref{2.8}) in terms of the functional
determinant of the Neumann Green's function (or the inverse of the
determinant of $F(\nabla)$ subject to Neumann boundary conditions).
Simultaneously, it gives the variational definition of these
functional determinants which specifies how the Neumann boundary
conditions enter them and how the action of differential operators
should be understood in the sense of integration by parts. The
derivation of the gaussian integral (\ref{Z1}) with fixed fields at
the boundary can be done along the same lines \cite{Feynman} and
leads to a similar variational definition with $G_D$ replacing
$G_N$.

\section{Heat kernel of the square-root type operator}
The heat kernel (\ref{4.34}) obviously satisfies the following
Dirichlet problem
 $K(\,\tau\,|\,x,y\,)$:
    \begin{eqnarray}
     &&(-\partial_{\tau}^2-\Box
     +m^2)\,K(\,\tau\,|\,x,y\,)=0, \\
     &&K(\,0\,|\,x,y\,)=\delta(x-y).   \label{B.1}
    \end{eqnarray}
Similarly to (\ref{2.19}) its solution is given by the
"brane-to-bulk propagator"
    \begin{eqnarray}
    K(\,\tau\,|\,x,y\,)
    = -G^{(\,d\!+\!1)}_D(\,x,\tau\,|\,y,\tau'\,)
    \stackrel{\leftarrow}{W}\!\big|_{\tau'=0},\,\,\,\,
    \stackrel{\leftarrow}{W}=
    -\stackrel{\leftarrow}{\partial}_{\tau'},  \label{B.2}
    \end{eqnarray}
where $G^{d\!+\!1}_D(\,x,\tau\,|\,y,\tau'\,)$ is the
$(d+1)$-dimensional Dirichlet Green's function of the operator
$\Box^{(d+1)}=\partial_\tau^2+\Box$ on half-space $\tau\geq 0$
    \begin{eqnarray}
     &&(m^2-\Box^{(d+1)})\;
     G^{(d\!+\!1)}_D(\,x,\tau\,|\,y,\tau'\,)=
     \delta(\tau-\tau')\,\delta(x,y), \\
     &&G^{(d\!+\!1)}_D(\,x,0\,|\,y,\tau'\,)=0.   \label{B.3}
    \end{eqnarray}
By the method of images one can construct it in terms of
$G^{(\,d\!+\!1\,)}(\,x,\tau\,|\,y,\tau'\,)$ -- the Green's function
in full space without boundary,
    \begin{eqnarray}
    G^{(\,d\!+\!1\,)}_D(\,x,\tau\,|\,y,\tau'\,)
    =G^{(\,d\!+\!1\,)}(\,x,\tau\,|\,y,\tau'\,)
    -G^{(\,d\!+\!1\,)}(\,x,\tau\,|\,y,-\tau'\,).    \label{B.4}
    \end{eqnarray}
For the massive case it reads
    \begin{eqnarray}
    G^{(\,d\!+\!1\,)}(\,x,\tau\,|\,y,\tau'\,)
    =\frac{1}{2\pi}
    \Big(\,\frac{m^2}{2\pi Z}\,\Big)^{\!\frac{d\!-\!1}{2}}
    K_{\frac{d\!-\!1}{2}}(Z),\,\,\,
    Z=m\sqrt{|\,x-y|^2+(\tau-\tau')^2}.     \label{B.5}
    \end{eqnarray}
Substituting (\ref{B.4})-(\ref{B.5}) into (\ref{B.2}) and using
recurrent relation between the modified Bessel functions of
different orders one obtains (\ref{4.35}).

\section*{Acknowledgements}
The work of A.O.B. on this paper was supported by the Russian
Foundation for Basic Research under the grant No 05-02-17661. The
work of D.V.N. was supported by the RFBR grant No 05-01-01049.
D.V.N. is also grateful to the Center of Science and Education of
the Lebedev Physics Institute and target funding program of the
Presidium of Russian Academy of Sciences. This work was also
supported by the LSS grant No 1578.2003.2.

\end{document}